\documentclass[sigconf]{acmart}

\usepackage{bbm}
\usepackage{balance}
\usepackage{cleveref}
\usepackage{multirow}
\usepackage{graphicx}
\usepackage{enumitem}
\usepackage[ruled, lined, linesnumbered, commentsnumbered, longend]{algorithm2e}
\usepackage{xcolor}

\newcommand{\squishlist}{
 \begin{list}{$\bullet$}
  { \setlength{\itemsep}{0pt}
     \setlength{\parsep}{3pt}
     \setlength{\topsep}{3pt}
     \setlength{\partopsep}{0pt}
     \setlength{\leftmargin}{1.5em}
     \setlength{\labelwidth}{1em}
     \setlength{\labelsep}{0.5em} } }

\newcommand{\squishlisttwo}{
 \begin{list}{$\bullet$}
  { \setlength{\itemsep}{0pt}
     \setlength{\parsep}{0pt}
    \setlength{\topsep}{0pt}
    \setlength{\partopsep}{0pt}
\setlength{\leftmargin}{2em}
\setlength{\labelwidth}{1.5em}
\setlength{\labelsep}{0.5em} } }

\newcommand{\squishend}{
\end{list}  }

\AtBeginDocument{%
  \providecommand\BibTeX{{%
    \normalfont B\kern-0.5em{\scshape i\kern-0.25em b}\kern-0.8em\TeX}}}

\copyrightyear{2023} 
\acmYear{2023} 
\setcopyright{acmlicensed}\acmConference[WWW '23]{Proceedings of the ACM
Web Conference 2023}{May 1--5, 2023}{Austin, TX, USA}
\acmBooktitle{Proceedings of the ACM Web Conference 2023 (WWW '23), May
1--5, 2023, Austin, TX, USA}
\acmPrice{15.00}
\acmDOI{10.1145/3543507.3583192}
\acmISBN{978-1-4503-9416-1/23/04}
    
\begin{document}

\title{Enhancing User Personalization in Conversational Recommenders}
\author{Allen Lin}
\affiliation{%
  \institution{Texas A\&M University}
  \city{College Station, Texas}
  \country{USA}}
\email{al001@tamu.edu}

\author{Ziwei Zhu}
\affiliation{%
  \institution{George Mason University}
  \city{Fairfax, Virginia}
  \country{USA}}
\email{zzhu20@gmu.edu}

\author{Jianling Wang}
\affiliation{%
  \institution{Texas A\&M University}
  \city{College Station, Texas}
  \country{USA}}
\email{jlwang@tamu.edu}

\author{James Caverlee}
\affiliation{%
  \institution{Texas A\&M University}
  \city{College Station, Texas}
  \country{USA}}
\email{caverlee@tamu.edu}

\begin{abstract}
Conversational recommenders are emerging as a powerful tool to personalize a user's recommendation experience. Through a back-and-forth dialogue, users can quickly hone in on just the right items. Many approaches to conversational recommendation, however, only partially explore the user preference space and make limiting assumptions about how user feedback can be best incorporated, resulting in long dialogues and poor recommendation performance. In this paper, we propose a novel conversational recommendation framework with two unique features: (i) a \textit{greedy NDCG attribute selector}, to enhance user personalization in the interactive preference elicitation process by prioritizing attributes that most effectively represent the actual preference space of the user; and (ii) a \textit{user representation refiner}, to effectively fuse together the user preferences collected from the interactive elicitation process to obtain a more personalized understanding of the user. Through extensive experiments on four frequently used datasets, we find the proposed framework not only outperforms all the state-of-the-art conversational recommenders (in terms of both recommendation performance and conversation efficiency), but also provides a more personalized experience for the user under the proposed multi-groundtruth multi-round conversational recommendation setting.

\end{abstract}

\ccsdesc[500]{Information systems~Recommender systems}

\keywords{Conversational Recommender System, User Personalization}

\maketitle
\section{Introduction}
Conversational recommender systems (CRSs) have attracted increasing attention in the research community due to their wide range of potential applications and promising functionality \cite{xu2021adapting, deng2021unified, ear, crm}. Rather than inferring from past user-item interactions, a CRS directly engages with the user through a conversation to learn the user's current and fine-grained preferences. Such a conversational approach can provide recommendations that are both highly personalized and justifiable. While several settings have been proposed to formulate the task of conversational recommendation, in this work, we focus on the \textbf{m}ulti-round \textbf{c}onversational \textbf{r}ecommendation (MCR) setting, which is widely perceived to be the most realistic setting so far \cite{systemask_user_respond, xu2021adapting, deng2021unified, ear, lin2022towards}. Under the MCR setting, a conversational recommender system starts with the \textit{interactive preference elicitation} process in which the system explores the preference space of the user, via asking the user's preferences on certain attributes (e.g., are you looking for \textit{country} music?), to obtain an accurate estimation of the user's current interests. Based on the learned preferences of the user, the system recommends the top-$k$ items that best match the user's current interests. 

\par The MCR setting has shown promising results across many recommendation tasks (e.g., music recommendations \cite{crm, xu2021adapting}, business recommendations \cite{scpr, ear}, and E-Commerce \cite{deng2021unified}), yet it makes a strict assumption -- that a user only has \textit{one target-item} in mind when interacting with the system. In reality, users often have diverse preferences which makes them simultaneously interested in a wide variety of items \cite{qiu2019user, qin2014contextual, tan2021sparse, zhang2022multiple}. Take music recommendations as an example, while Alice likes the \textit{pop} style song `Shape of You', she also finds a \textit{country} style song -- such as `Somebody like you' -- enjoyable to listen. In this situation, designing and evaluating a CRS with such a strong single-groundtruth assumption greatly limits the generalizability of the system.        
\par Hence, we extend the MCR setting to the \textbf{m}ulti-\textbf{g}roundtruth \textbf{m}ulti-round \textbf{c}onversational \textbf{r}ecommendation (MGMCR) setting, under which the CRS is designed and evaluated with the realistic formulation that a user could be simultaneously interested in a variety of items. This MGMCR setting is more challenging than the traditional MCR setting, as the CRS now aims to discover all the groundtruth items of the user. The key in achieving promising recommendation performance lies in the effectiveness of \textit{user personalization} in the system. Specifically, we identify two key challenges that existing works face under the MGMCR setting: \\
(1) \textit{lack of user personalization in interactive preference elicitation}: During the interactive preference elicitation process, a CRS aims to explore a user's preference space by asking the user's preference on a set of domain specific attributes. Since every user has \textit{unique} preferences, a CRS must prioritize on asking the attributes that the user would consider relevant in expressing her actual preference space. Although there are some differences among how existing CRSs decide on which attribute(s) to prioritize for a specific user, most of them rely heavily on the use of \textit{entropy}, which does not provide sufficient granularity for the system to identify the most relevant attribute(s) for the user. Such an issue could even be exacerbated under the MGMCR setting given the user now has a broader preference space. In turn, the asked attributes often result in a sub-optimal representation of the user's actual preference space, resulting in poor recommendation performance. \\    
(2) \textit{lack of user personalization in preference fusion}: To achieve personalized recommendations for the user, a CRS must fuse together the user's preferences -- collected in the interactive preference elicitation process -- to form a personalized understanding of the user. To do that, many existing works \cite{deng2021unified, xu2021adapting, scpr, ear} have adopted the use of \textit{attribute-filtering} in which the system treats every item that does not align perfectly with the collected preferences as a mismatch for the user. This approach has two limitations under the MGMCR setting. First, many groundtruth items could be mistakenly considered as mismatches. For instance, Alice is interested in both the \textit{action} movie, Top Gun: Maverick, and the \textit{comedy} movie, Despicable Me. So when the system asks Alice if she is looking for a \textit{comedy} movie, Alice will respond with a confirmation. However, with the attribute \textit{comedy} added to her set of collected preference, Top Gun: Maverick would now be considered as a mismatch for Alice since it does not contain the attribute \textit{comedy}. Second, it does not take the user's past interests into consideration, overlooking the scenario where the same preference could indicate different signals for different users. For a user who has rarely watched \textit{horror} movies, a confirmation on the attribute \textit{horror} could indicate a stronger shift in preference compared to a user who has frequently watched \textit{horror} movies.  

\par To address the aforementioned challenges, we propose a novel framework for enhancing \textbf{u}ser \textbf{p}er\textbf{s}onalization in conversational \textbf{rec}ommendation systems (UpsRec). To enhance user personalization in the interactive preference elicitation process, the framework adopts an NDCG-oriented greedy attribute selection strategy in which the system directly prioritizes on asking the attributes that most effectively personalize the recommended items for each user. And to enhance user personalization in fusing the collected user preferences, the framework deploys a user representation refiner to iteratively refine the initial user representation learned from past interactions. With the refined representation being a more accurate reflection of the user's current interests, items no longer need to be filtered to ease the difficulty of ranking, avoiding the scenario of mistakenly excluding groundtruth items. In summary, the main contributions of this work are as follows:
\begin{itemize}[leftmargin=.35cm]
    \item We are the first to comprehensively study the MGMCR scenario for conversational recommendation, which is more realistic and generalizable than the existing MCR setting.  
    \item We propose the UpsRec framework with unique features designed to provide personalized experiences for the users and effectively guide them to their items of interest.
    \item We show the proposed UpsRec framework significantly outperforms state-of-the-art CRS methods by an average of 23\%, while requiring fewer interactions with the user.   
\end{itemize}

\section{Preliminaries}
\label{sec:prelim}
In this section, we provide a brief introduction on \textit{multi-round conversational recommendation} (MCR), which is the most widely adopted setting for conversational recommendation \cite{ear, xu2021adapting, deng2021unified, scpr}. 
\par Formally, let \begin{math} V \end{math} denote the itemset, and \begin{math} P = (p_0, p_1,..., p_m) \end{math} denote a set of $m$ domain-specific attributes that describe an item \begin{math} v \in V \end{math}. During each conversation session, a CRS aims to first learn the user's fine-grained preferences on the asked attributes, then provides personalized recommendations. A conversation session starts by the user specifying an attribute \begin{math} p_0 \end{math} contained in the target item(s) that she is asking the system to find (e.g., Can you find me some \textit{country} music?). Next, a CRS calls upon its conversation component to decide whether to ask the user more attributes or make a recommendation. If the policy agent decides not enough preference evidence has been collected, it will pick one attribute \begin{math} p \end{math} from the set of unasked attributes \begin{math} P_{cand} \end{math} to prompt the user. If the user likes the asked attribute, then policy agent updates the set of collected preferences of the user, \begin{math} P_{u} \end{math}, by adding \begin{math} p \end{math} to \begin{math} P_u \end{math}; otherwise \begin{math} P_u  \end{math} remains unchanged. If the policy agent decides enough information has been collected after $t$ turns of interaction, the CRS then calls upon its recommender component to make a list of recommendations. The recommender component would then rank all the items in the candidate itemset \begin{math} V_{cand} \end{math} -- the itemset that contains only items with attributes perfectly matching all specified preferences in \begin{math} P_u \end{math} -- then recommends the top-K ranked items to display to the user. If the user accepts the recommendation, then the system quits. If the user rejects all the recommended items, the system first removes the rejected item \begin{math} V_{rej} \end{math} from \begin{math} V_{cand} \end{math} then repeats the cycle by calling upon its conversational component to decide the next action to take. This process continues until the user quits due to impatience or a predefined maximum number of turns has been reached.   

\smallskip
\noindent\textbf{Multi-round Multi-Groundtruth Setting.} Although MCR is currently perceived as one of the most realistic settings for conversational recommendation, it makes the limiting assumption that a user only has one item of interest when interacting with the system. Therefore, we propose the new multi-round multi-groundtruth (MGMCR) setting, which makes the following changes: (1). when being asked by the system on the preference towards an attribute $p$, the user is now assumed to answer according to the attributes associated with \textit{all} her groundtruth items (i.e., can be one or multiple), which more realistically mimics how a user would interact with a CRS; and (2). the system now only generates one conversation session for each user $u$, instead of a new conversation session for each observed user-item interaction pair of the user.

\section{Proposed Framework}
In this section, we introduce a novel framework for enhancing \textbf{u}ser \textbf{p}er\textbf{s}onalization in conversational \textbf{rec}ommendation systems, namely UpsRec. UpsRec is designed to tackle two key challenges: (i) how to bring user personalization to interactive preference elicitation; and (ii) how to bring user personalization to preference fusion. Specifically, we design a \textit{greedy NDCG attribute selector} that personalizes the interactive preference elicitation process by utilizing an NDCG-oriented attribute selection strategy to prioritize on attributes that most effectively represent the actual preference space of the user. Then, to effectively fuse together the user preferences collected from the interactive preference elicitation process, a \textit{user representation refiner} is deployed to iteratively refine the current representation of the user.

\par The framework starts with the \textit{latent interest estimation} stage where the system builds an initial predictive model to estimate the interest of users by learning a set of latent representations from the historical user-item interactions. Then a reinforcement learning based policy agent, integrated with both the \textit{greedy NDCG attribute selector} and the \textit{user representation refiner}, is deployed to decide how to interact with the user. Together, the entire flow of the proposed UpsRec framework is detailed in Algorithm \ref{alg:upsflow}.

\begin{algorithm}
    \caption{UpsRec}
    \label{alg:upsflow}
    \SetKwInOut{KwIn}{Input}
    \SetKwInOut{KwOut}{Output}
    \KwIn{the user-item interaction matrix \textbf{M}; the item-attribute information Dictionary $\textbf{D}_p$; user $u$; a matrix containing the adjacent attributes of all users $\textbf{P}_{adj}$ (see \Cref{sec:greedy_process} for details); all items $V$; the number of items to recommend $k$; the maximum number of turns $T$}
    \KwOut{k recommended items}
    Set: $\textbf{R}_{U}$, $\textbf{R}_{V}$, $\textbf{R}_{P}$ = \textit{Latent Interest 
    Estimation}\ ($\textbf{M}$, $\textbf{D}_p$) \\
    Initialize: $\textbf{r}_{u_0}$ = $\textbf{R}_{U}[u]$; $\textbf{r}_{u_t}$ = $\textbf{R}_{U}[u]$ \\
    Initialize: $P_u$ = $\{\}$; $P_{cand}$ = $ \textbf{P}_{adj}[u]$; \ $V_{cand} = V$ \\
    \For{turn $t = 1,2,3 ... T$}{
        \textit{Policy Agent} selects an action $a$ \newline
        \uIf{$a == a_{ask}$}{ 
           $p_{ask}$ = \textit{Greedy NDCG Attribute Selector}\ ($\textbf{r}_{u_0}$, $\textbf{r}_{u_t}$, $P_u$, \newline $P_{cand}$) \newline
           Update: $P_{cand}$ = $P_{cand}$ $\backslash$ \ $\textit{p}_{ask}$ \newline
           \uIf{$u$ {$\normalfont$ accepts} $p_{ask}$}{
                Update: $P_u = P_u \cup \ \textit{p}_{ask}$\newline
                Set: $\textbf{r}_{u_t}$ = \textit{User Representation Refiner} ($\textbf{r}_{u_0}$, $P_u$)
           }
        }
        \Else{
            Rank all items in $V_{cand}$ \\
            Recommends the top $k$ ranked items, $V_{rec}$, to $u$; \\
            \uIf{$u$ {\normalfont accepts} $V_{rec}$}{
                Return $V_{rec}$; Exit.
            }
            \Else{
                Update $: V_{cand} = V_{cand} \backslash \ V_{rec}$
            }
        }
    }    
\end{algorithm}

\subsection{Latent Interest Estimation}
\label{sec:latentlearn}
The goal of the \textit{Latent Interest Estimation} stage is to learn an initial set of latent representations for all users, items, and attributes -- denoted respectively as $\textbf{R}_U, \textbf{R}_V, \textbf{R}_P$ -- that will later be used throughout the framework. Following \cite{scpr, ear, xian2019reinforcement}, we adopt a factorization machine \cite{rendle2010factorization} (FM) as the predictive model. Given user $u$, the user's collected preferences $P_u$, and an item $v$, $u$'s predicted interest in $v$ is computed as:
\begin{equation}
    \label{eq:y_score}
    y(u, v, P_u) = \textbf{r}_u^T\textbf{r}_v + \sum_{p_i \in P_u} \textbf{r}_v^T\textbf{r}_{p_i}
\end{equation}
where $\textbf{r}_u$ and $\textbf{r}_v$ respectively denote the latent representation of user $u$ and item $v$, and $\textbf{r}_{p_i}$ denotes the latent representation for attribute $p_i \in P_u$. As shown in \cite{lin2022quantifying}, re-weighting an item's relevance based on its frequency in the training set results in more effective learned latent representations. Therefore, we modify the standard pairwise Bayesian Personalized Ranking (BPR) objective function \cite{bprloss} to a weighted variation as follows: 
\begin{equation}
    \begin{aligned}
        L = \frac{1}{e^{n_1f_i}}[-ln\sigma(y(u,v,P_u)) - \sigma(y(u,v',P_u))] + e^{n_2f_i}||\textbf{v}||^2 \\
        + \lambda_{\hat{\theta}}(||\hat{\theta}||^2)
    \end{aligned}
    \label{equ:w_bpr}
\end{equation}
where $e$ denotes the natural log; \begin{math} f_i \end{math} denotes the interaction frequency of the item \begin{math} v \end{math}; \begin{math} \frac{1}{e^{n_1f_i}}\end{math} controls the relevance of the item $v$ to the loss function; and \begin{math} e^{n_2f_i}\end{math} controls the scale of the regularization on the item's learned latent representation $\textbf{v}$. And \begin{math} n_1 \end{math} and \begin{math} n_2 \end{math} denote hyperparameters that control the impact of the interaction frequency on $v$'s relevance and regularization scale, and \begin{math} \lambda_{\theta} \end{math} denotes the regularization parameter to prevent overfitting.

\subsection{Greedy NDCG Attribute Selector}
\label{sec:greedy_process}
Once the conversation begins, how do we \textit{personalize} the elicitation of a user's preferences? To achieve this, we propose a greedy NDCG attribute selector which utilizes an NDCG-oriented attribute selection strategy to help the CRS identify the attributes that most \textit{effectively} represent the current preference space of the user. Since the eventual goal of a CRS is to make accurate recommendations for the user, we consider the \textit{effectiveness} of an attribute equivalent to how much it can increase the recommendation performance. That is, after being asked by the system, how much can knowing the user's preference on that particular attribute contribute to the refinement of the user's recommendations? To this end, the NDCG-oriented greedy attribute selection strategy identifies the attribute with the highest expected increase in recommendation performance (measured in NDCG), from the user's unasked \textbf{adjacent} attributes \begin{math} P_{cand} \end{math}, to ask the user. The workflow of the proposed greedy NDCG attribute selector is illustrated in Algorithm \ref{alg:as_flow}.  

\begin{algorithm}
    \SetKwInOut{KwIn}{Input}
    \SetKwInOut{KwOut}{Output}
    \KwIn{the initial representation of the user $\textbf{r}_{u_0}$; the current representation of the user at the $t$-th turn $\textbf{r}_{u_t}$; \ collected preferences of the user $P_{u}$; \ the unasked adjacent attributes of the user $P_{cand}$}
    \KwOut{the attribute to ask next $p_{ask}$}
    Initialize: $p_{ask}$ = \ $\phi$; $max\_ndcg = 0$ \\
   
    \For{$p$ in $P_{cand}$}{
        Set: $\hat{P}_{u}$ = $P_u$ $\cup$  $p$ \\
        Refine the initial user representation with $\hat{P}_{u}$:\\ ${\textbf{r}}_{u_{t+1}}$ =  \textit{User Representation Refiner}\ ($\textbf{r}_{u_0}$, $\hat{P}_u$) \\
        Set: $p\_ndcg$ = \textbf{NDCG}(${\textbf{r}}_{u_{t+1}}$) - \textbf{NDCG}($\textbf{r}_{u_t}$) \\
        \uIf{$p\_ndcg$ > $max\_ndcg$}{ 
            $max\_ndcg$ = $p\_ndcg$ \\
            $p_{ask}$ = $p$ \\
        }
    }
    \KwRet{$p_{ask}$}
    \caption{Greedy NDCG Attribute Selector}
    \label{alg:as_flow}
\end{algorithm}

It is worth noting $P_{cand}$ contains only the unasked \textbf{adjacent attributes} of the user -- attributes that are contained in the user's interacted items and have not been asked before -- which significantly reduces the search space  \cite{scpr}. To start, the attribute selector takes in the initial representation of the user learned from past interactions $\textbf{r}_{u_0}$, the current representation of the user $\textbf{r}_{u_t}$, the set of collected preferences of the user $P_{u}$, and the set of candidate attributes $P_{cand}$ as inputs. Then for every unasked candidate attribute $p$ in $P_{cand}$, we compute its expected NDCG gain. First, we add $p$ to $P_{u}$ to simulate the scenario that the asked attribute, $p$, will be liked by the user. Then based on the updated $P_{u}$, denoted as $\hat{P}_{u}$, we use the user representation refiner to compute the refined representation $\textbf{r}_{u_{t+1}}$ which reflects the system's understanding of the user after knowing the user's preference on $p$. Then, we use $\textbf{r}_{u_{t+1}}$ and $\textbf{r}_{u_{t}}$ to compute the NDCGs of the user, both before and after knowing the user's preference on $p$, and use the difference to serve as the expected NDCG gain for $p$. Once we have computed the expected NDCG gains for all the candidate attributes in $P_{cand}$, we pick the attribute with the highest expected NDCG gain to be asked next, $p_{ask}$. It is important to note that the NDCG of a user is computed on the \textbf{validation set}, instead of the \textbf{testing set}. Specifically, we first rank all the items that are not included the training samples of the user by taking the dot product of the item's representation (learned in \Cref{sec:latentlearn}) with the \textit{input user representation}. Then we use the ranked items of the user to compute the DCG and the groundtruth items of the user (in the \textbf{validation set}) to compute the Ideal DCG, which together give us the NDCG of the user.   

\subsection{User Representation Refiner}
After eliciting a user's preferences, a natural next question is how do we \textit{personalize} the system's model of the user? To provide a personalized fusion for the collected preferences from the interactive preference elicitation process, we deploy a user representation refiner to refine the initial latent representation of the user, \begin{math} \textbf{r}_{u_0} \end{math}, to be a more accurate reflection of the user's current interests, by considering the collected preferences of the user, \begin{math} P_u \end{math}. Formally, we have:
\begin{displaymath}
    \mathcal{F}(\textbf{r}_{u_0}, P_u) = \hat{\textbf{r}}_{u}
\end{displaymath}
where \begin{math} \mathcal{F} \end{math} denotes the user representation refiner which maps the initial representation of the user, \begin{math} \textbf{r}_{u_0} \end{math}, to the refined representation, \begin{math} \hat{\textbf{r}}_{u} \end{math}, based on the collected preferences of the user, \begin{math} P_u \end{math}. 

\smallskip
\noindent\textbf{Input Layer:}
To fuse the collected preferences in a personalized manner, the user representation refiner takes both the \textit{past} and the \textit{current} interests of the user, respectively encoded in $\textbf{r}_{u_0}$ and $P_{cand}$, into consideration. First to construct the inputs, we fetch, from $\textbf{R}_P$ (learned in \Cref{sec:latentlearn}), the latent representations of all specified preferences in \begin{math} P_u \end{math}, denoted as \begin{math} \{\textbf{r}_{p_i} | 1 \leq i \leq |P_u|\}\end{math}, and stack them together into a matrix \begin{math} \textbf{R}_{p_u} \in \mathbb{R}^{|P_u|\times d} \end{math}, where \begin{math} d \end{math} is the dimensionality of the latent representation.  

\smallskip
\noindent\textbf{Preference Aggregation Layer:}
Then to account for the contextual consistency of the neighboring preferences, an attention-based preference aggregation layer is applied to \begin{math} \textbf{R}_{p_u} \end{math}. Following \cite{vaswani2017attention}, we have: 
\begin{displaymath}
    \hat{\textbf{R}}_{p_u} = softmax(\frac{\textbf{R}_{p_u} \textbf{W}^q \: \textbf{R}_{p_u}^T \textbf{W}^k}{\sqrt{d}}) \: \textbf{R}_{p_u}\textbf{W}^v
\end{displaymath}
where \begin{math} \hat{\textbf{R}}_{p_u} \in \mathbb{R}^{|P_u| \times d} \end{math} denotes a set of preference representations refined by incorporating the contextual information of the neighboring preferences. And \begin{math} \textbf{W}^q \end{math}, \begin{math} \textbf{W}^k \end{math}, and \begin{math} \textbf{W}^v \end{math} denote the projection matrices. Next, we apply the weighted-sum to aggregate all the refined preference representations to construct a new representation that comprehensively represents the collected preferences of the user as an entity:
\begin{displaymath}
    \textbf{z}_{p_u} = \sum_{i \in |P_u|} \alpha_{i} \cdot \hat{\textbf{r}}_{p_i} 
\end{displaymath}
where \begin{math} \textbf{z}_{p_u} \end{math} denotes the aggregated representation that comprehensively represents all the specified preferences in $P_u$ as an entity; $\hat{\textbf{r}}_{p_i}$ denotes a particular updated preference representation in $\hat{\textbf{R}}_{p_u}$. And \begin{math} \alpha_i \end{math} is the weight learned by the vanilla attention network: 
\begin{displaymath}
    \alpha_i = \textbf{r}_{u_0}^T \ tanh(\textbf{W}^c\textbf{r}_{p_i} + \textbf{b}_1) \quad
    \alpha_i = \frac{exp(\alpha_i)}{\sum_{n\in M} exp(\alpha_n)}
\end{displaymath}
where \begin{math} \alpha_i \end{math} is a scalar that represents the relevance of the representation of a particular attribute, \begin{math} \hat{\textbf{r}}_{p_i} \end{math}, \begin{math} \textbf{W}^c \end{math} is the projection matrix, and \begin{math} \textbf{b}_1 \end{math} is the bias vector. Note that the initial latent representation of the user, \begin{math} \textbf{r}_{u_0} \end{math}, is directly used as the `query' in calculation of \begin{math} \alpha_i \end{math} to enhance personalization in the calculation of the relevance of a particular attribute, $\alpha_i$, to the user. In this way, \begin{math} \alpha_i \end{math} is learned by taking into account -- (1) the intrinsic properties of individual attributes; (2) the contextual consistency of the neighboring attributes; and (3) the user's general past interests -- so that \begin{math} \textbf{z}_{p_u} \end{math} is a comprehensive representation of $P_u$. 

\smallskip
\noindent\textbf{Update Layer:} Once \begin{math} \textbf{z}_{p_u} \end{math} has been computed, we concatenate it with the user's initial latent representation, \begin{math} \textbf{r}_{u_0} \end{math}, to obtain a concatenated representation \begin{math} \textbf{r}_{u_{cat}} \end{math} that contains information of the user's past and current interests:
\begin{displaymath}
    \textbf{r}_{u_{cat}} = \textbf{z}_{p_u} \oplus \textbf{r}_{u_0}
\end{displaymath}
Then, a one-layer feed-forward neural network is applied to obtain the refined representation of the user \begin{math} \hat{\textbf{r}}_{u}\end{math}:
\begin{displaymath}
    \hat{\textbf{r}}_{u} = \textbf{W}\textbf{r}_{u_{cat}} + \textbf{b}_2
\end{displaymath}
where \begin{math} \textbf{W} \end{math} is the weight matrix of the neural network and \begin{math} \textbf{b}_2 \end{math} is the bias vector.

\smallskip
\noindent\textbf{Objective Function:} Since the intention of the user representation refiner is to enhance personalization in fusing the collected preferences of the user ($P_u$), the refined representation of the user, \begin{math} \hat{\textbf{r}}_u \end{math}, should have higher predicted affinity scores (calculated using dot-product) with items that align well with \begin{math} P_u \end{math}. To this end, we adopt the standard pairwise Bayesian Personalized Ranking (BPR) objective function for optimizing the user representation refiner as a personalized ranking algorithm. Note we did not use the weighted variation of the BPR loss, \Cref{equ:w_bpr}, since it is not applicable in this situation. Formally, we have:
\begin{displaymath}
    L_{bpr} = \sum_{(v,v', P_u) \in  \mathcal{D}} -ln\sigma(\hat{y}(u, v, P_u) - \hat{y}(u, v', P_u)) + \lambda_{\theta} ||\theta||^2 
\end{displaymath}
with \begin{math} \hat{y}(u, v, P_u) \end{math} is computed as:
\begin{equation}
    \label{eq:yhat}
    \hat{y}(u, v, P_u) = \mathcal{F}(\textbf{r}_{u_0}, P_u) ^T  \: \textbf{v}
\end{equation}
Here, \begin{math} \mathcal{D} \end{math} is a set of pairwise instances for the BPR training, \begin{math} u \end{math} is the user whose embedding is being updated, \begin{math} \sigma \end{math} is the sigmoid function, \begin{math} \lambda_{\theta} \end{math} is the regularization parameter to prevent overfitting, \begin{math} \textbf{r}_{u_0} \end{math} is the initial user representation, and \begin{math} \mathcal{F} \end{math} is the user representation refiner. Note, \Cref{eq:yhat} involves the use of $\mathcal{F}$ in its computation, which allows $\mathcal{F}$ to be trained through back-propagation. In addition, \Cref{eq:yhat} drops the second item-attribute affinity term used in \Cref{eq:y_score}. This is because the attribute information has already been fused into the refined user presentation, $\hat{\textbf{r}}_u$, making the item-attribute affinity term redundant.        

\smallskip
\noindent\textbf{Pairwise Instances Sampling:}
In this section, we introduce the pairwise instances sampling strategy used to construct \begin{math} \mathcal{D} \end{math}, for training the user representation refiner. For each user \begin{math} u \end{math}, we randomly sample one of \begin{math} u \end{math}'s interacted items to serve the positive sample, \begin{math} v \end{math}. Then, we randomly keep \begin{math} n \end{math}, where \begin{math} T-1 \geq n \geq 1 \end{math} (T being the maximum allowed turns of interaction), attribute(s) from the \begin{math} v \end{math}'s list of attributes to serve as \begin{math} P_u \end{math}. Lastly, we randomly sample one of \begin{math} u \end{math}'s non-interacted items, whose attributes share a low \textit{Jaccard Similarity} with \begin{math} P_u \end{math}, to serve as the negative sample, \begin{math} v' \end{math}. Compared to the standard pairwise instance sampling where \begin{math} P_u \end{math} would be \begin{math} v \end{math}'s entire list of attributes, we randomly select n attributes to keep the user representation refiner robust to a wide variety of attribute combinations. Note, \begin{math} v' \end{math} is specifically chosen to be an item that share few attribute(s) with \begin{math} v \end{math} to emphasize the importance of \begin{math} P_u \end{math}.

\subsection{Action Selection Policy Learning}
Following \cite{scpr}, we adopt a two-layer feed forward neural network as the policy network and use the standard Deep Q-Learning \cite{mnih2015human} for optimization. As shown by previous works \cite{deng2021unified, scpr}, massive action space hinders both the training and the performance of RL-based policy networks; thus, we define our action space to be $\mathcal{A} = \{a_{ask}, \: a_{rec}\}$. During each interaction with the user, the policy network takes in the state vector $\textbf{s}_t$ and outputs the Q-values for both $a_{ask}$ and $a_{rec}$ to be their estimated rewards. The CRS will then choose the action with higher reward and interact with the user accordingly. The state vector $\textbf{s}_t$ is a concatenation of two vectors:
\begin{displaymath}
    \textbf{s}_t = \textbf{r}_{u_0} \oplus \textbf{s}_t^{his}
\end{displaymath}
where $\textbf{r}_{u_0}$ denotes the initial user representation, which is expected to add personalization into the policy network, and $\textbf{s}_t^{his}$ denotes the conversation history up until turn $t$, which is expected to guide the system to act more wisely -- e.g., if multiple asked attributes have been accepted, the system might be ready to recommend. Extending from \cite{scpr}, we define the five kinds of rewards used for training the policy network: (1) $r_{rec\_suc}$, a NDCG-oriented positive reward when recommendation succeeds; (2) $r_{rec\_fail}$, a negative reward when the user rejects the recommendation;  (3) $r_{ask\_suc}$, a slightly positive reward when the user confirms the asked attribute; and (4) $r_{ask\_fail}$, a negative reward when the user rejects the asked attribute, and (5). $r_{stop}$, a strongly negative reward when reaching the maximum number of turns. 

\section{Experiments}
In this section, we showcase experiments over multiple datasets to evaluate the performance of the proposed UpsRec approach to answer three key research questions: \textbf{RQ1}. How does the recommendation performance of the proposed framework compare to existing CRS approaches? \textbf{RQ2}. Does the proposed \textit{greedy NDCG attribute selector} enhance user personalization in the interactive preference elicitation process? \textbf{RQ3}. Does the proposed \textit{user representation refiner} enhance user personalization in the preference fusion process? 

\subsection{Experiments Setup}
\label{sec:setup}
\smallskip
\noindent\textbf{Dataset.} We evaluate the proposed framework on four benchmark datasets that are widely adopted to evaluate multi-round conversational recommender systems \cite{crm, ear, scpr, xu2021adapting, deng2021unified}. 
\squishlist
  \item \textbf{Lastfm} and \textbf{Yelp}. Lastfm is a dataset used to evaluate music artist recommendation and Yelp is used to evaluate business recommendation. In \cite{ear}, Lei et al. manually categorize the original 8,432 attributes in Lastfm into 33 coarse-grained categories, and construct a 2-layer taxonomy to reduce the original 590 attributes to 29 first-layer categories for Yelp.

  \item \textbf{Lastfm*\footnote{https://grouplens.org/datasets/hetrec-2011/}} and \textbf{Yelp*\footnote{https://www.yelp.com/dataset/}}. In \cite{scpr}, Lei et al. propose it is unrealistic to manually merge attributes for practical applications, so they adopt the original attributes for these two datasets. For a fair comparison, we conduct experiments on both versions.
\squishend
  All datasets contain only the implicit (binary) feedback from the user (e.g. whether a user has interacted with the item or not). Following previous works \cite{he2017neural, ear}, we keep users with at least 10 interactions to alleviate data sparsity. Statistics of all the datasets are summarized in Table~\ref{tab:table1}. 
\begin{table}
  \begin{center}
    \caption{Dataset statistics}
    \label{tab:table1}
    \begin{tabular}{l r r r r}
      \hline
      Dataset & \#users & \#items & \#interactions & \#attributes \\

      Lastfm & 1,801 & 7,432 & 76,693 & \textbf{33} \\

      Lastfm* & 1,801 & 7,432 & 76,693 & \textbf{8,432} \\

      Yelp & 27,675 & 70,311 & 1,368,606 & \textbf{29} \\

      Yelp* & 27,675 & 70,311 & 1,368,606 & \textbf{590} \\ 
      \hline
    \end{tabular}
  \end{center}
\end{table}

\smallskip
\noindent\textbf{Baselines.}
To examine the proposed framework, we compare its performance with the following start-of-the-art baseline conversational recommendation approaches.     
\squishlist
  \item \textbf{Matrix Factorization (MF)}: MF is a collaborative filtering algorithm that is widely used for many \textit{static} recommendation tasks. We report its performance to demonstrate the general effectiveness of CRSs.
  \item \textbf{Max Entropy (MaxEnt)}: MaxEnt is a rule-based attribute selection strategy. At each turn, the policy agent selects the attribute with the highest entropy to ask. Recommendation is made either when the candidate space is small enough, or the policy agent runs out of attributes.
  \item \textbf{CRM \cite{crm}}: CRM is a CRS that uses a belief tracker to record a user's preference conveyed, and trains a policy network via reinforcement learning to decide how to interact with the user. We follow \cite{ear} to adapt it to the multi-round conversational setting for fair comparison. 
  \item \textbf{EAR \cite{ear}}: Similar to CRM, EAR also learns a predictive model to estimate a user's preference and trains a policy network to determine whether to ask more attributes or make recommendations. In addition, it also considers the feedback from the user to further fine-tune the learned predictive model.
  \item \textbf{SCPR \cite{scpr}}: Extending from EAR, SCPR leverages the concept of adjacent attributes to construct a knowledge graph to reduce the search space of attributes and uses Deep Q-Learning \cite{mnih2015human} to learn a more efficient policy network.
  \item \textbf{UNICORN \cite{deng2021unified}}: Unicorn is the current state-of-the-art CRS. It applies a dynamic-weighted-graph based RL approach to integrate the conversation and the recommendation components and deploys an weighted-entropy action selection strategy to reduce the candidate action space. 
\squishend
Since all baselines are initially proposed under the MCR setting, for a fair comparison, we retrain the conversational components of all baseline models to adapt to the proposed MGMCR setting as introduced in \Cref{sec:prelim} and adopt a NDCG-oriented reward function to optimize the policy networks of all baseline models.   

\smallskip
\noindent\textbf{Metrics.} To evaluate a CRS under the MGMCR setting, we adopt the Top-K metrics: Hit Rate (HT@K) and Normalized Discounted Cumulative Gain (NDCG@K). HT@K could be considered as an extension of the widely adopted \cite{crm, deng2021unified, ear, xu2021adapting} Success Rate (SR) metric under the MCR setting. Instead of measuring if the one groundtruth item exists in the K recommended items, HT@K measures what fraction of the K recommended items are of the groundtruth items. To provide a more comprehensive evaluation, NDCG@K is also adopted to account for the rank order in recommendations. In addition, we also report Average Turns (AT) to evaluate the efficiency of the system. If the conversation session reaches the predefined maximum turn T, then the turn count for the session is T.         

\smallskip
\noindent\textbf{Training Details.} 
Following \cite{ear}, we split the user-item interactions in the ratio of 7:2:1 for training, validation, and testing, and set the size of the recommendation list $k$ to 10 and the maximum turn $T$ to 15. The training process is divided into offline and online stages. The offline training is intended for building: (i) the predictive model (FM) for estimating the latent interests of the user; and  (ii) the user representation refiner for incorporating the collected preferences to iteratively refine the representation of the user. Afterwards, we follow the user simulation process, introduced in \Cref{sec:prelim}, to conduct online training to optimize the policy agent (network) on how to best interact with the user via reinforcement learning. We tune all hyper-parameters on the validation set and set them as follows: the dimensionality, $d$, of all the user, item, attribute representations are set to 64. SGD is used for optimizing the weighted BPR loss adopted in the predictive model used in \Cref{sec:latentlearn}, with $n_1$ set to 7, $n_2$ set to 8, $\lambda$ set to .001. Adam is used for optimizing the standard BPR loss adopted in the user representation refiner, with $\lambda$ set to .002. For constructing the pairwise instance $\mathcal{D}$, we collect 15,000 pairwise samples for each user with the Jaccard Similarity threshold set to 0.33. For training the policy network, we set the experience replay memory size to 50,000, the sample batch size to 256, and the discount factor $\gamma$ to 0.95.

\begin{table*}[h]
  \begin{center}
    \caption{Overall Recommendation Performance Comparison, with NDCG@10 $\uparrow$, HT@10 $\uparrow$, and AT $\downarrow$ used as evaluation metrics. Boldface indicates  statistically significant improvement ($\rho$ < 0.01) over all baselines.}
    \label{tab:table2}
    \setlength{\tabcolsep}{5pt}
    \renewcommand{\arraystretch}{1}
    \scalebox{.97}{
        \begin{tabular}{c c c c c c c c c c c c c} 
          \hline
          \multicolumn{1}{c}{ } & \multicolumn{3}{c}{\textbf{Lastfm}} & \multicolumn{3}{c}
          {\textbf{Lastfm*}} & \multicolumn{3}{c}{\textbf{Yelp}} & \multicolumn{3}{c}{\textbf{Yelp*}}\\ 
          \hline
          \multicolumn{1}{c}{} & NDCG@10 & HT@10 & AT & NDCG@10 & HT@10 & AT & NDCG@10 & HT@10 & AT & NDCG@10 & HT@10 & AT\\
          \hline
          MF & .0415 & .0684 & NA & .0392 & .0679 & NA & .0199 & .0419 & NA & .0165 & .0305 & NA \\
         
          MaxEnt  & .1033 & .1289 & 12.80 & .1819 & .2253 & 8.36 & .2668 & .2870 & 8.44 & .1694 & .2041 & 13.10\\
          
          CRM  & .1359 & .2006 & 11.13 & .1710 & .2237 & 7.60 & .2422 & .2920 & 7.24 & .1072 & .1289 & 13.69\\
          
          EAR  & .1455 & .2111 & 10.75 & .1765 & .2283 & 7.55 & .2503 & .3012 & 7.11 & .1095 & .1307 & 13.71 \\ 
         
          SCPR & .1671 & .2407 & 9.94 & .2079 & .2752 & 6.90 & .2529 & .3016 & 7.09 & .3031 & .2924 & \underline{12.77}\\
          
          Unicorn &  \underline{.1982} & \underline{.2808} & \underline{8.95} & \underline{.2239} & \underline{.2818} & \underline{6.84} & \underline{.2557} & \underline{.3041} & \underline{6.99} &  \underline{.3118} & \underline{.2977} & 12.92\\
          \hline
          \hline
          UpsRec  & \textbf{.3144} & \textbf{.3358} & \textbf{5.99} & \textbf{.3278} & \textbf{.3437} & \textbf{5.89} & \textbf{.3631} & \textbf{.3805} & \textbf{5.72} & \textbf{.3853} & \textbf{.3754} & \textbf{6.57}\\
          \hline
        \end{tabular}
    }
  \end{center}
\end{table*}

\subsection{Performance Comparison (RQ1)}
As shown in \Cref{tab:table2}, UpsRec significantly outperforms all the baseline models in both recommendation performance (measured in both NDCG@10 and HT@10) and conversation efficiency (measured in AT). This validates our hypothesis that enhancing user personalization, in both the interactive preference elicitation process and the preference incorporation process, is an effective strategy to build a CRS. We also make the following observations: \\
(1). All conversational recommendation approaches significantly outperform the static matrix factorization (MF) model, showing the promising functionality of CRSs in general. \\
(2). For all baseline models, we observe a significant gap between the two recommendation performance metrics, especially on the Lastfm and the Lastfm* datasets. In general, the measured NDCG@10 is significantly lower than the measured HT@10. This is because NDCG@10 is measured by taking the rank order of the groundtruth items into account. Since baseline models only rely on \textit{attribute-filtering} to fuse the collected preferences of the user, the user representation -- used for ranking the candidate items to calculate NDCG -- is often not an accurate reflection of the user's current preference space, since it has never been updated after learned from the past user-item interactions. As a result, even when the groundtruth items can be successfully recommended, they are usually ranked low on the recommendation list. Compared to the baseline models, UpsRec show similar recommendation performance measured in both NDCG@10 and HT@10. This is because UpsRec integrates the \textit{user representation refiner} to iteratively update a user's representation whenever a new preference has been collected. In turn, the user representation used for ranking the candidate items will be a more realistic reflection of the user's current interests, resulting in groundtruth items being ranked higher on the recommendation list when they are successfully recommended. \\       
(3). Comparing to all other baseline CRSs, UpsRec shows consistently higher conversational efficiency (measured by AT) and recommendation performance (measured by both NDCG@10 and HT@10) across all four tested datasets, especially on the Yelp* dataset. This indicates that UpsRec can scale up to both large attribute spaces (8,432 attributes in Lastfm* v.s. 29 attributes in Yelp) and larger number of user-item interactions (1,368,606 interactions in Yelp* v.s. 76,693 interactions in Lastfm), which is critical for deploying CRSs in practice.

\begin{figure}
      \includegraphics[width=.4591\textwidth]{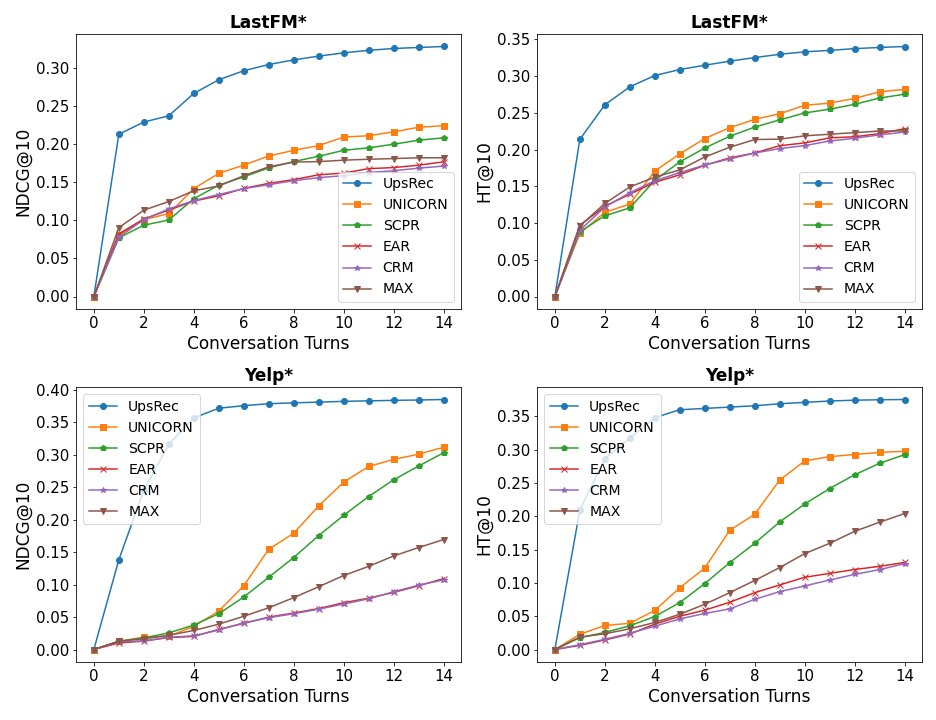}
  \caption{CRS Performance vs Conversation Turns}
  \Description{}
  \label{fig:ndcg_hit_by_turn}
\end{figure}

\begin{figure}
  \includegraphics[width=.47\textwidth]{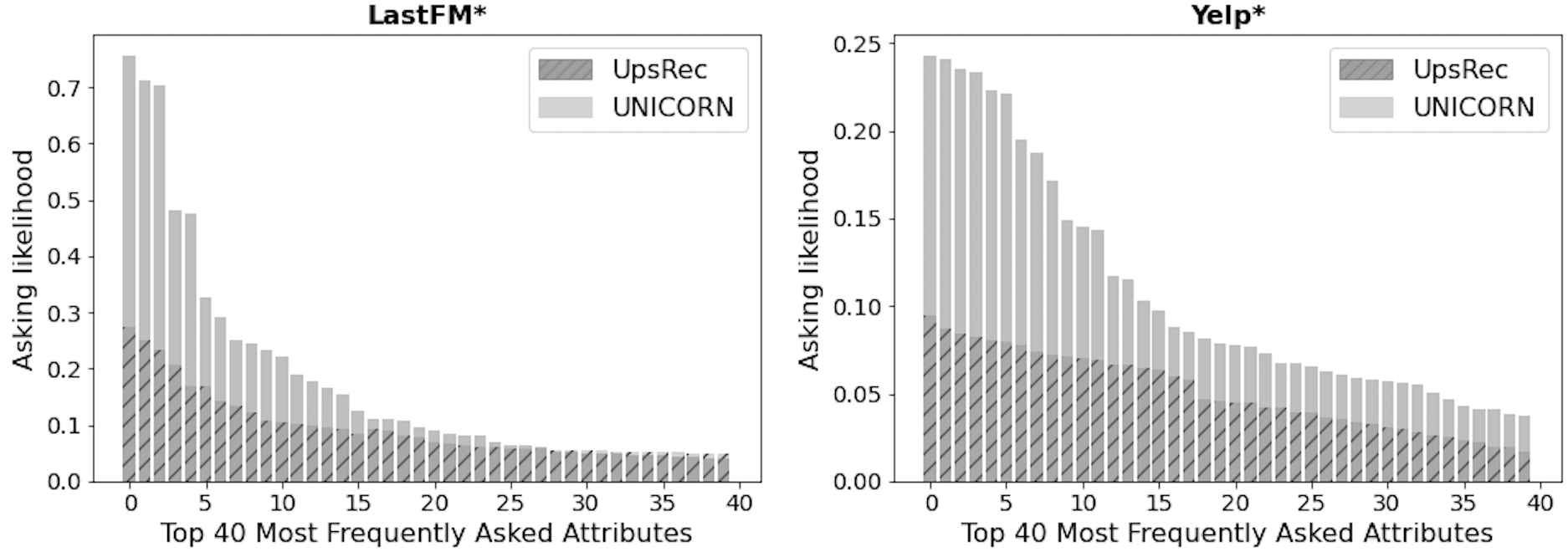}
  \caption{Asking Likelihood of the Top 40 Most Frequently Asked Attributes in UpsRec and UNICORN}
  \label{fig:asked_attr_freq}
\end{figure}

\subsection{Enhancing User Personalization in Interactive Preference Elicitation (RQ2)}
To determine whether the proposed \textit{greedy NDCG attribute selector} enhances user personalization in the interactive preference elicitation process, we examine both the \textit{effectiveness} and the \textit{efficiency} of the asked attributes in representing user's actual preference space.

\par
\begin{figure}
\vspace{-2mm}
  \includegraphics[width=.435\textwidth]{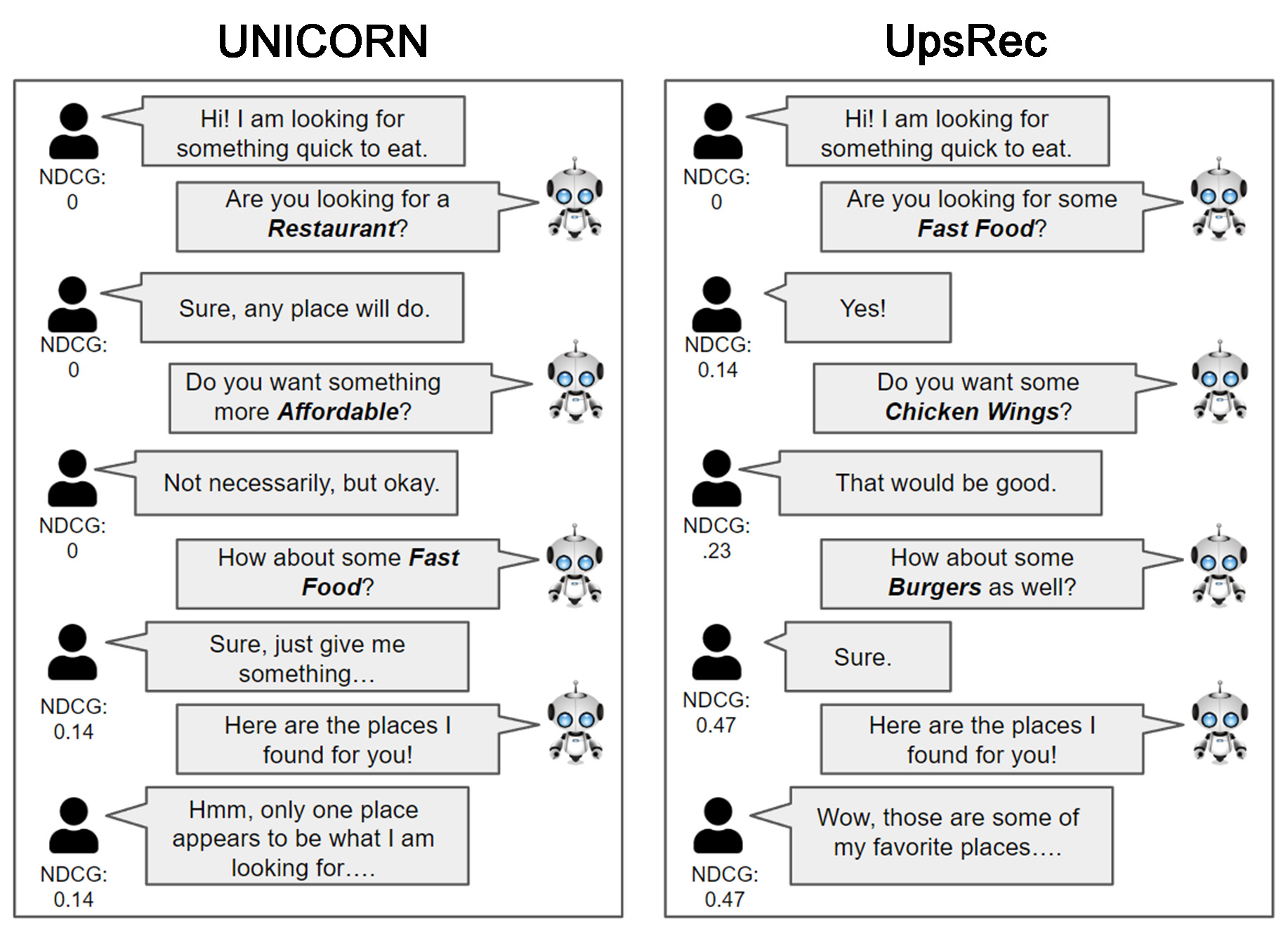}
  \caption{Sample Conversations by UNICORN and UpsRec}
  \Description{}
  \label{fig:sample_conv}
\end{figure}

 Since the goal of a CRS is to make highly personalized recommendations, we consider the effectiveness of an asked attribute equivalent to how much it can increase the recommendation performance. As shown in \Cref{fig:ndcg_hit_by_turn}, knowing the user's preferences on the attributes -- selected by the proposed greedy NDCG attribute selector -- brings significantly higher increments in recommendation performance than knowing the user's preferences on the attributes selected by all the baseline CRSs. This indicates that the proposed greedy NDCG attribute selector selects attributes that are more relevant to the user, such that knowing the user's preferences on these attributes helps the system gain a more comprehensive understanding of the user's current interests. Next, we consider the efficiency of the attributes selected by the greedy NDCG attribute selector. From \Cref{fig:ndcg_hit_by_turn}, for UpsRec, we observe a strong diminishing increase in recommendation performance after asking five attributes. This indicates that the asked attributes are highly efficient in representing the user's actual preference space, such that knowing the user's preferences on four asked attributes helps the system to achieve roughly 80\% of its best recommendation performance on the Lastfm* dataset or nearly 90\% on the Yelp* dataset. Compared to UpsRec, all other baselines show marginal increases in recommendation performance after knowing the user's preference on the first few asked attributes (\Cref{fig:ndcg_hit_by_turn}). This is due to the lack of user personalization in the interactive preference elicitation. Compared to the proposed NDCG-oriented attribute selection strategy, the entropy-based attribute selection strategy adopted by existing works cannot precisely identify the set of attributes that most efficiently represent the user's actual preference space. To better illustrate, we compare the averaged asking likelihood of the top-40 most frequently asked attributes in both UNICORN and UpsRec (\Cref{fig:asked_attr_freq}). The asking likelihood of an attribute can be interpreted as the percentage of users that the system predicts will find the asked attribute relevant. For instance, the top-three most frequently asked attributes in UNICORN have an average asking likelihood of 0.7. That is, UNICORN thinks that 70\% of the users will consider these three attributes important in expressing their preferences. However, due to the lack of user personalization in the weighted-entropy attribute selection, such prediction is often inaccurate. As can be observed in \Cref{fig:asked_attr_freq}, UNICORN exhibits a higher asking likelihood for almost all the top-40 most frequently asked attributes, on both the LastFM* and the Yelp* dataset (\Cref{tab:table2}). This indicates that many of the asked attributes in UNICORN are inefficient in representing the user's actual preference space. A sample conversation of such, on the Yelp* business recommendation dataset, is shown in \Cref{fig:sample_conv}. While the attributes \textit{Restaurant} and \textit{affordable} have high calculated weighted-entropy, knowing the user preference on them has virtually no contribution to the recommendation performance of the system. Compared to the weighted-entropy attribute selection strategy used in UNICORN, the proposed NDCG-oriented attribute selection strategy identifies attributes that more efficiently represent the preference space of the user.

\subsection{Enhancing User Personalization in Fusing Preferences (RQ3)}
\begin{figure}
  \includegraphics[width=.431\textwidth]{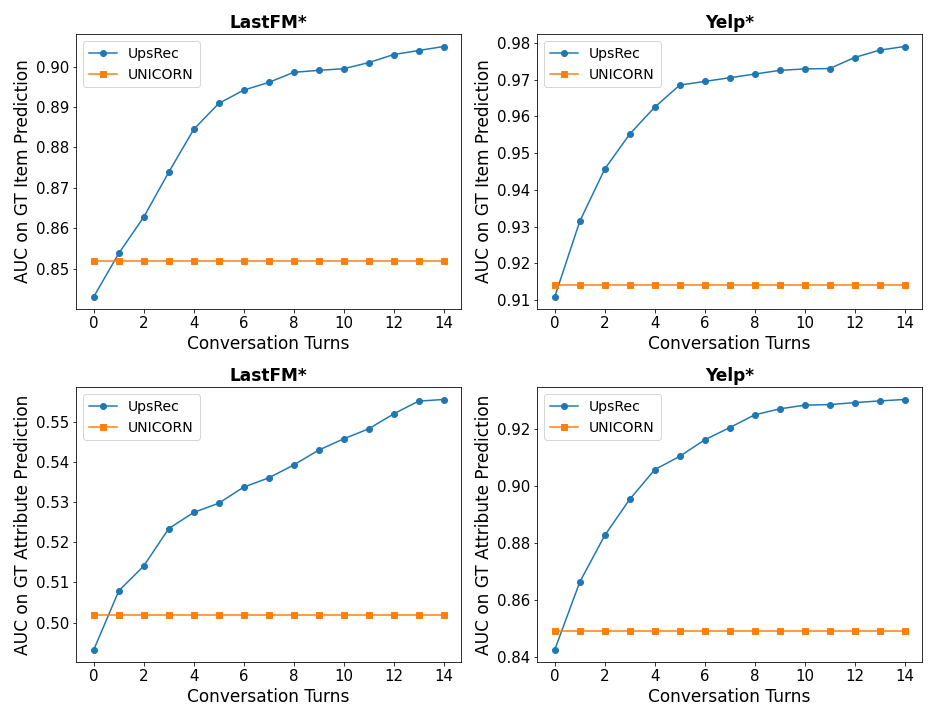}
  \caption{AUC Scores on GroundTruth(GT) Item Prediction and GT Attribute Prediction at Different Conversation Turns}
  \label{fig:relative_auc}
\end{figure}

A personalized preference fusing process aims to consistently increase the system's understanding of the user by providing a personalized modeling based on the user's collected preferences. Therefore, to determine the effectiveness of the proposed \textit{user representation refiner} in enhancing user personalization in the collected-preference incorporation process, we examine whether the refined representation of the user more precisely reflects the user's actual items of interest and attributes of interest, compared to the current state-of-the-art CRS -- UNICORN. To examine if the refined user representation more precisely reflects the user's actual items of interest, we compare the AUC Scores on groundtruth item prediction (computed following \cite{bprloss}) between UpsRec and UNICORN. As shown in \Cref{fig:relative_auc}, by iteratively refining the representation of the user once a new preference of the user has been collected, UpsRec's AUC gradually increases as the conversation goes on. Note UNICORN adopts a different strategy for learning the initial user representations which results in different AUC scores at turn 0; however, since UNICORN (and all baselines) performs no refinement on the learned initial user representations, its AUC remains constant and is surpassed by UpsRec's in one turn. This indicates that compared to the attribute-filtering approach used in UNICORN, the proposed user representation refiner is able to help the system obtain a uniquely \textit{personalized} modeling of the user, which more precisely reflects the user's items of interests. A similar pattern is also observed in (\Cref{fig:relative_auc}). As the conversation goes on, UpsRec's AUC on groundtruth attributes prediction also gradually surpasses the AUC of UNICORN. Note, the AUC score on groundtruth attributes prediction is computed by treating all the attributes that exist in any of the groundtruth items as the positive attributes and the attributes that do not as the negative attributes. A higher AUC score indicates that the proposed user representation refiner is able to help the system establish a more \textit{personalized} modeling of the user, which facilitates the process of identifying the attributes that the user would find relevant in expressing her preference space. 

\section{Related Work}
Traditional recommender systems challenges of (i) how to capture a user's precise preferences; and (ii) how to provide a human-interpretable justification for their recommendations \cite{jannach2021survey, gao2021advances}. Many existing works have approached these challenges with auxiliary data to better interpret user intentions \cite{xian2019reinforcement, he2015trirank, cheng2019mmalfm, zhang2020explainable, zhang2014explicit}. In another direction, the emergence of conversational recommenders that dynamically interact with users through real-time interactions are able to elicit current and detailed preferences directly from users \cite{jannach2021survey}. In this way, a CRS can precisely learn a user's interests along with a series of actions that provide justification evidence.
\par In general, a CRS is composed of two components -- the conversation component and the recommendation component. The conversation component decides which action (further collecting user preference or calling upon the recommender component to make recommendations) to take during each interaction with the user. The recommendation component learns a set of user, item, and attribute embeddings for ranking all the candidate items. While early works on CRSs primarily rely on template or choice-based conversation components for collecting user preferences \cite{10.1145/2792838.2799681, 10.1287/opre.2014.1292, zou2020neural}, recent works have explored the possibilities of more effective preference elicitation methods and conversational strategies with the help of reinforcement learning. For example, CRM \cite{crm} proposes a deep reinforcement-based conversation component that builds a personalized preference elicitation process. Inspired by CRM, EAR \cite{ear} proposes a novel optimization function to enhance the richness of the embeddings learned by the recommendation component, then designs a three-stage framework that further strengthens the interaction between the recommender component and the conversation component. While sharing the same recommender component, SCPR \cite{scpr} integrates knowledge graphs to improve the reasoning ability of the conversation component. Further utilizing the informativeness of knowledge graphs, UNICORN \cite{deng2021unified} proposes a dynamic weighted graph based reinforcement approach that systematically integrate the recommendation and the conversation components. However, the above-introduced approaches \cite{crm, ear, scpr, deng2021unified, xu2021adapting} all adopt an entropy-based attribute-selection strategy which limits user personalization in the preference elicitation process. In addition, these approaches only make coarse uses of the obtained user preferences, refraining the system from gaining a deeper understanding of the user. In this work, we propose a novel attribute-selection and a preference fusion strategy that jointly address these two issues. 
Note that while there are other research directions in CRSs including dialogue understanding \cite{zheng2020pre, jin2018explicit, li2016deep, zou2020responses,wang2022learning}, response generation \cite{zhou2020improving, lin2004rouge, liu2020towards, papineni2002bleu}, and exploration-exploitation trade-offs \cite{zhang2020conversational, yu2019visual, zhou2020conversational, lee2019melu, iovine2022empirical} those are not the focus of this work.

\section{Conclusion}
In this work, we propose a novel framework, named UpsRec, with two unique features aiming to enhance user personalization in both the interactive preference elicitation process and the preference fusion process. The experimental results on four frequently adopted conversational recommendation datasets show that the proposed framework not only outperforms all baseline CRS methods but also brings a more personalized experience for the user under the more realistic and generalizable MGMCR setting.

\bibliographystyle{ACM-Reference-Format}
\balance
\bibliography{ref}

\end{document}